\documentclass[reprint,aps,prb]{revtex4-1}

\usepackage{graphicx}
\usepackage{dcolumn}
\usepackage{bm}
\usepackage{color}
\usepackage{natbib}

\begin{document}

\title{Intrinsic and extrinsic spin-orbit coupling and spin relaxation in monolayer PtSe$_2$}

\author{Marcin Kurpas}
\email{marcin.kurpas@us.edu.pl}
\affiliation{Institute of Physics,University of Silesia in Katowice, 41-500 Chorz\'{o}w, Poland}

\author{Jaroslav Fabian}
\affiliation{Institute for Theoretical Physics, University of Regensburg, Regensburg 93040, Germany}

\date{\today}

\begin{abstract}
Monolayer PtSe$_2$ is a semiconducting transition metal dichalcogenide characterized by an indirect band gap, space inversion symmetry, and high carrier mobility. Strong intrinsic spin-orbit coupling and the possibility to induce extrinsic spin-orbit fields by gating make PtSe$_2$ attractive for fundamental spin transport studies as well as for potential spintronics applications. We perform a systematic theoretical study of the spin-orbit coupling and spin relaxation in this material. Specifically, we employ first principles methods to obtain the 
basic orbital and spin-orbital properties of PtSe$_2$, also in the presence of an external transverse electric field. We 
calculate the spin mixing parameters $b^2$ and the spin-orbit fields $\Omega$ for the Bloch states of electrons and holes. This information
allows us to predict the spin lifetimes due to the Elliott-Yafet and D'yakonov-Perel mechanisms. We find that $b^2$ is rather large, 
on the order of $10^{-2}$ and $10^{-1}$, while $\Omega$ varies strongly
with doping, being about $10^{3} - 10^{4}$\,ns$^{-1}$ for 
carrier density in the interval  $10^{13}-10^{14}$\,cm$^{-2}$
at the electric field of 
1 V/nm. We estimate the spin lifetimes to be on the picosecond level. 
\end{abstract}

\pacs{Valid PACS appear here}
\maketitle

%
%
\section{\label{sec:level1}Introduction}
Transition metal dichalcogenides (TMDCs) have been 
investigated---mainly in the bulk form but also as layered slabs---for many decades \cite{Frindt1966,Wilson1969,Mattheiss1973,Kam1982,Joensen1986}. The recent revival of interest in TMDCs has been fueled by a broad range of fascinating electronic, optical, and spin properties of two-dimensional (2D) samples of TMDCs, which are stable in air. The possibility of 
controlling physical properties of TMDCs by, e.g., stacking \citep{Lebegue2009,Mak2010,Kuc2011,Splendiani2010,Ciarrocchi2018,Avsar2019}, doping \citep{Mouri2013}, straining \citep{Ma2012,Zollner2019}, 
or gating \citep{Ross2013}, demonstrates their potential for electronic \citep{Radisavljevic2011},  optoelectronic \citep{Britnell1311} and valleytronic \citep{Langer2018} applications. 
Moreover, due to strong spin-orbit coupling and the presence of a semiconducting gap, TMDCs are also well suited for applications in spintronics \citep{Zutic2004:RMP,Fabian2010}, as they can induce
spin-orbit coupling (SOC) into graphene via strong proximity effect
\citep{Avsar2014,Gmitra2015}.

\begin{figure}
    \centering
    \includegraphics[width=0.85\columnwidth]{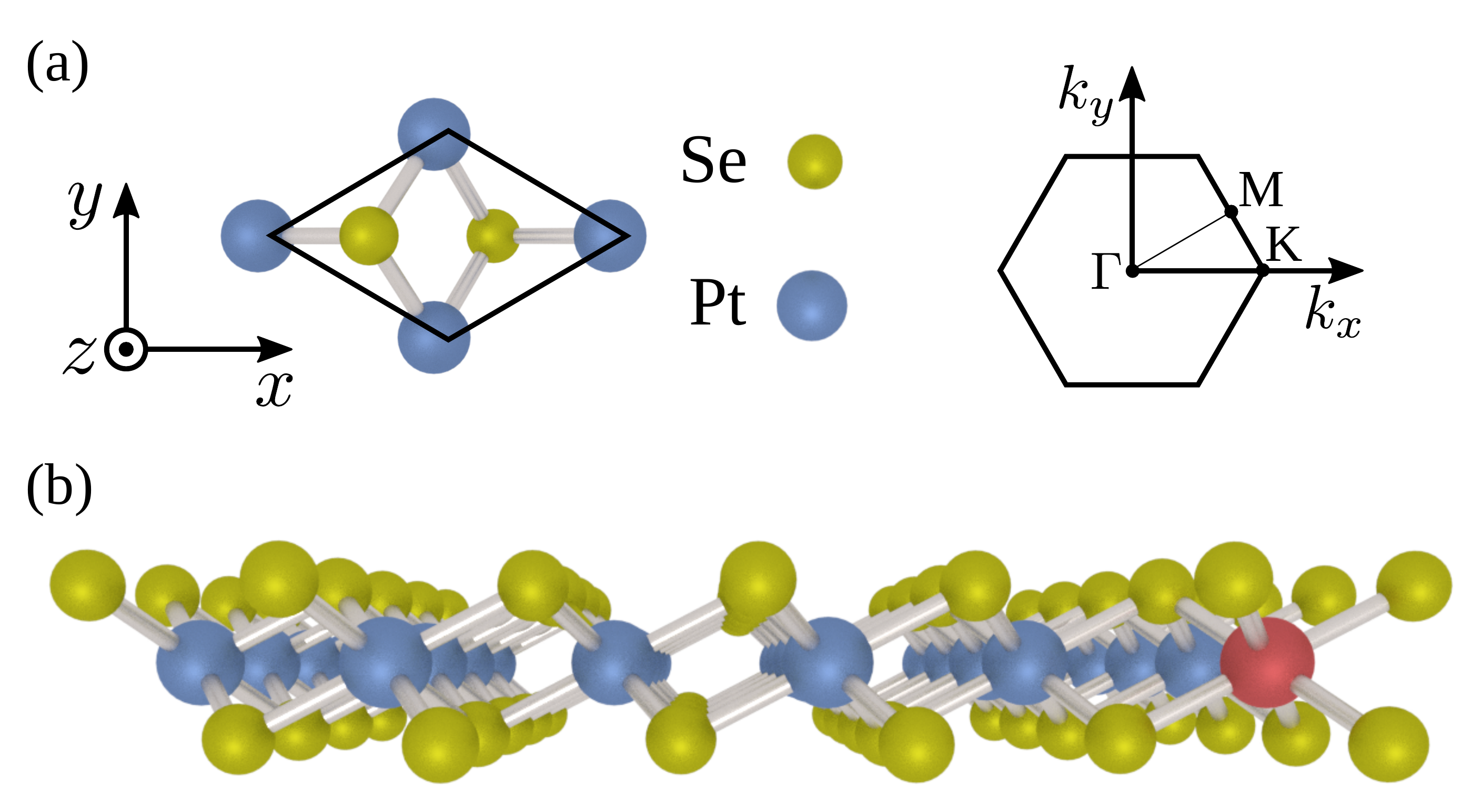}
    \caption{\label{fig:structure} Sketch of the crystalline structure of monolayer PtSe$_2$. (a) Top view on the unit cell and the corresponding first Brillouin zone with indicated high symmetry points. (b) Side view of the atomic structure of monolayer PtSe$_2$. The center of inversion is at the Pt atom, marked red.
    }
\end{figure}

Recently demonstrated atomically thin PtSe$_{2}$ \citep{wang2015,Yu2018} is a distinct member of the 2D TMDC family. What sharply distinguishes this material from other TMDCs is its high room-temperature carrier mobility \citep{Zhao2017}, which is close to that of phosphorene  \citep{Avsar2017}. 
But in contrast to phosphorene,  PtSe$_{2}$ exhibits good stability when exposed to air \citep{Zhao2017}.
Like other TMDCs, a monolayer of PtSe$_2$ consists of an atomically thin layer of transition metal (Pt) within two layers of chalcogen (Se) atoms [Fig. \ref{fig:structure} (a),(b)]. It crystallizes in the centrosymmetric structure of $P3\bar{m}1$ space group being  isomorphic with the  $D_{3d}$ point group. 
While bulk PtSe2 is metallic, in the monolayer limit it is a semiconductor with a sizeable indirect gap reported to be in the
range of $1.2-2$\,eV \citep{Huang2016,Zhang2017,Zhao2017,Ciarrocchi2018,Kandemir2018}. \\
Monolayer PtSe$_{2}$  also holds promise to exhibit rich spin phenomena. One of the most exciting is the hidden spin polarization \citep{Zhang2014} of degenerate bands near the Fermi level, recently observed in ARPES experiments \citep{Yao2017}. Its origin is attributed to local site dipole fields (local Rashba effect) generating opposite helical spin textures for spin degenerate states, spatially resolved with respect to different Se layers. The opposite dipole fields compensate each other leaving the total crystal potential inversion symmetric, and thus preserving the spin degeneracy of bands \citep{Zhang2014}.  Another interesting phenomenon is the defect induced magnetism, which is reported for mono \citep{Zhang_2016,Zulfiqar2016} and multilayer \citep{Avsar2019} PtSe$_{2}$ slabs. In the latter case, a magnentic phase can be switched between ferro- to antiferromagnetic by changing the parity of the number of layers \citep{Avsar2019}. The interplay of such 
magnetic effects with spin-orbit coupling could lead to 
interesting magnetotransport phenomena. 

Strong spin-orbit coupling, intrinsic band gap, and high carrier mobility make PtSe$_2$ a good candidate for building spintronic devices, such as, a spin valve or spin transistor \citep{DDas1990}. Essential for these devices is a coherent (ensemble) dynamics of the electron spin. Such a dynamics is disrupted by spin dephasing and spin relaxation processes. Thus, the question concerning the electron spin lifetime in monolayer PtSe$_2$ is of great importance for potential applications of this material in spintronics. This question has not yet been 
systematically addressed theoretically.

Here, we investigate the problem of the spin relaxation
in monolayer PtSe$_2$ by employing first principles calculations
and extracting useful information about the spin-orbit coupling and spin relaxation. Two mechanisms dominating spin relaxation in non-magnetic materials, such as PtSe$_2$, are considered. Namely, the Elliott-Yafet \citep{elliott_theory_1954,Yafet_1963} and D'yakonov-Perel \citep{dyakonov1971R,dyakonov1971}.
In the Elliott-Yafet mechanism, the intrinsic SOC mixes opposite spin components of degenerate Bloch states. In effect, an electron can  flip its spin upon momentum scattering, with the probability given by the so called \textit{spin mixing} parameter $b^2_{\mathbf{k}}$. It is related to spin relaxation time $\tau_{s,\rm EY}$ via the formula \citep{elliott_theory_1954,Monod1979,fabian1998}
\begin{equation}    
   \tau_{s,\rm EY}^{-1}\approx 4 b^2 \tau_p^{-1},
   \label{eq:elliott}
\end{equation}
where $\tau_p^{-1}$ is the momentum relaxation rate, and $b^2$ is the Fermi surface average of $b_{k}^2$. \\
In the D'yakonov-Perel mechanism spins randomize their phase via the interaction with the fluctuating Rashba fields $\Omega_{k}$. These fields appear due to broken space inversion symmetry of the structure, e.g.,  due to a substrate or an external electric field. The initial phase of spins is completely randomized after the time  $\tau_{s,\rm DP}$
\begin{equation}
     \tau_{s,\rm DP}^{-1} =   \Omega^2_{\perp}    \tau_p,
     \label{eq:tau_dp}
 \end{equation}
 where $\Omega^2_{\perp}$ denotes the Fermi surface average of the squared spin-orbit field component perpendicular to the spin orientation  $\Omega^2_{{\mathbf k}\perp}$.
In realistic systems these two mechanisms usually coexist and compete with each other. 
Here we show that for both mechanisms 
the spin relaxation in PtSe$_2$ is very fast, up to a few picoseconds for experimentally accessible momentum scattering time. Thus, PtSe$_2$ does not appear to be the best material for building spintronic devices requiring long spin lifetimes. However, it should be useful for investigating spin-orbit induced transport phenomena.  

The paper is organized as follows.  In Section \ref{sec:methods} we briefly describe methods and details of calculations.  Section  \ref{sec:results} contains results of our first principles calculations with a discussion, including  effects of the intrinsic and extrinsic SOC on the band structure, spin mixing parameter, and spin-orbit fields. Estimations of spin lifetime due to  Elliott-Yafet and D'yakonov-Perel relaxation mechanisms are also included here. Section \ref{sec:conclusions} contains final conclusions.

%
\section{Methods}\label{sec:methods}
First principle calculations were performed using {\sc Quantum Espresso} package \citep{QE-2009,QE-2017}. The norm--conserving pseudopotential with the  Perdew-Burke-Ernzerhof (PBE) \citep{perdew_1996,*perdew_1997}  version of the generalized gradient approximation (GGA) exchange--correlation potentials was used.
The kinetic energy cutoff of the plane wave basis sets was $50\,$Ry  for the wave function and $200$\,Ry  for charge density. These values were found to give converged results also for spin related quantities. 
Self consistency was achieved with $21\times 21\times 1$ Monkhorst-Pack grid while for structure optimization a smaller grid  $10 \times 10\times 1$ was chosen.
The initial lattice constant of PtSe$_2$ was taken from experiment \citep{wang2015} and was later optimized for the chosen pseudopotential using the variable cell and quasi-Newton schemes as implemented in the  {\sc Quantum ESPRESSO} package. During optimization process all atoms were free to move in all directions to minimize the internal forces below the threshold $10^{-4}$\,Ry/bohr. The calculated lattice constant is $a=3.748$\,\AA, very close to the experimental value in bulk 3.73\,\AA \citep{wang2015}, and is in a good agreement with other calculations \citep{Zhang_2016,Kandemir2018}.

The Fermi contour  averages of spin mixing parameter $b^2$ and spin-orbit field $\Omega^2$ entering the formulae (\ref{eq:elliott}) and (\ref{eq:tau_dp}) are  calculated using the formula
\begin{equation}
    A = \frac{1}{\rho(E_F)S_{BZ}}\int_{FC}\frac{A_{\mathbf{k}}}{\hbar |\mathbf{v}_F(\mathbf{k})|}dk,
\end{equation}
where $A_{\mathbf{k}}$ stands for $b^2_{\mathbf{k}}$ or $\Omega^2_{{\mathbf k}\perp}$,  $S_{BZ}$ is the area of the Fermi surface, $\rho(E_F)$ is the density of states per spin at the Fermi level, $\mathbf{v}_F(\mathbf{k})$ is the Fermi velocity and the integration takes over an iso-energy contour.
%
\section{Results and discussion}\label{sec:results}
\begin{figure}
    \centering
    \includegraphics[width=\columnwidth]{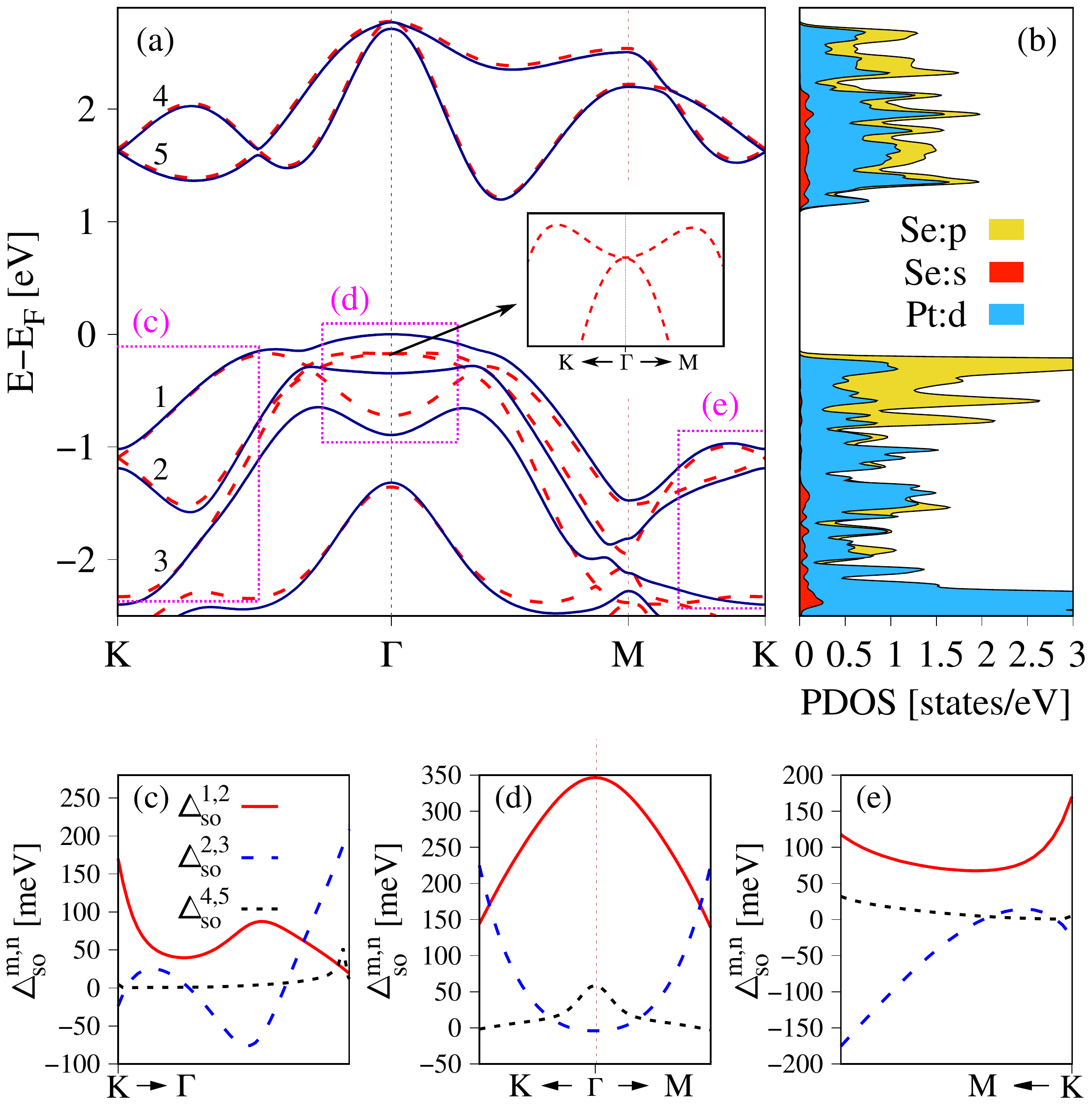}
    \caption{\label{fig:bs} Calculated non-relativistic (dashed line) and relativistic (solid line) band structures along high symmetry lines in the FBZ (a). The non-relativistic band structure is misaligned with the Fermi energy for better transparency. The inset shows a zoom of bands close to the $\Gamma$ point. Purple rectangles depict the range of $k$ points in (c)-(e). (b) Density of states projected onto atomic orbitals. (c)-(e) Extracted spin-orbital splittings $\Delta_{\text{so}}^{m,n}$ between the valence bands $m$ and $n$, labeled 1,2,3, and 4,5 in (a), calculated as a difference between interband energy distances with SOC and without SOC. }
\end{figure}

We first examine the orbital effects. The calculated non-relativistic and relativistic band structures are shown in Fig. \ref{fig:bs} (a).  Monolayer PtSe$_2$ is an indirect gap semiconductor with a sizeable band gap.
Without SOC the calculated band gap is $1.38\,$eV. The valence band (VB) maximum  is located slightly away (0.15\,\AA$^{-1}$) from the BZ center, while the conduction band (CB) minimum lies in the middle of the $\Gamma$M path. The band edge at the $\Gamma$ point is a saddle point lying 38\,meV below the global VB maximum [see the inset in Fig. \ref{fig:bs} (a)]. 
The valence and conduction bands close to the band gap are formed mainly by $d$-electrons of platinum and $p$-electrons of selenium [Fig. \ref{fig:bs} (b)]. In the valence band up to ~1\,eV below the Fermi level the dominant contribution comes from Se $p$-electrons, with significant admixture of $d$-electrons from Pt. In the conduction band the contributions from Pt and Se atoms are almost equal. 

\subsection{Intrinsic spin-orbit coupling.}
\textit{Spin-orbit splitting}.
Relativistic effects in PtSe$_{2}$ are significant. Spin-orbit coupling splits the originally two-fold (four-fold with spin) degenerate valence band at the $\Gamma$ point
into two (doubly spin degenerate) bands which are separated
by the spin-orbit split-off gap of $\Delta_{\text{so}}=350$\,meV.
As a result the maximum of the VB moves to the BZ center and the indirect band gap reduces to 1.2\,eV, in agreement with earlier calculations \citep{Zhao2017,Kandemir2018}.
The orbital degeneracy is also removed at the K-point. The energy splitting of the two highest valence bands [bands 1 and 2 in Fig. \ref{fig:bs} (a)] is 170\,meV. In the conduction band, the corresponding spin-orbital gaps $\Delta_{\text{so}}$ are much smaller, 59\,meV at the $\Gamma$ point and 5\,meV at the K-point.

Away from high symmetry points, we calculate 
the energy shifts $\Delta_{\text{so}}^{n,m}(\mathbf{k})=\Delta^{n,m}_{\text{rel}}(\mathbf{k}) - \Delta^{n,m}_{\text{nrel}}(\mathbf{k})$, where $\Delta^{n,m}_{\text{rel}(\text{nrel})}(\mathbf{k})$ is the energy difference between the bands $n$ and $m$ obtained from the relativistic (non-relativistic) calculation. 
It provides information about the shift of the bands upon turning on SOC,  with respect to their initial energy. This can be partially translated into the strength of the direct
spin-orbit interaction between bands $m$ and $n$, relative to the total SOC in the band $m$ coming from all possible couplings.  Considering that SOC leads to band repulsion, 
positive $\Delta_{\text{so}}^{n,m}$ means that the direct 
spin-orbit interaction between the bands $n$ and $m$ is likely dominant (with respect to couplings to other bands). Analogously, if $\Delta_{\text{so}}^{n,m}$ is negative, the spin-orbit  interaction between the bands $n$ and $m$ is weak enough to be overcome by couplings to others. Note that $\Delta_{\text{so}}^{n,m}=0$ does not mean  $\langle \psi_n|H_{so}|\psi_m\rangle=0$. Rather, it says that the direct SOC between bands $m$ and $n$ is of the same order as their couplings to the other bands, and no change in energy is observed.

The results for three valence bands and two conduction bands labeled  in Fig. \ref{fig:bs} (a) respectively  1, 2, 3 and 4, 5,  are shown in Fig. \ref{fig:bs} (c)-(e). We have checked, by tracing the irreducible representations of the bands and applying the group theory methods, that for all  $\Delta_{\text{so}}^{n,m}$s shown in Fig. \ref{fig:bs} (c)-(e) the direct SOC between bands $n$ and $m$ is allowed by the symmetry. 
For the valence bands 1-3, $\Delta_{\text{so}}^{n,m}(\mathbf{k})$ is strongly momentum dependent and takes significantly larger values than $\Delta_{\text{so}}^{4,5}$ in the conduction band. In the presented $k$-points range it varies from -180\,meV for $\Delta_{\text{so}}^{2,3}(\mathbf{k})$ [Fig. \ref{fig:bs} (e)] up to ~350\,meV for $\Delta_{\text{so}}^{1,2}(\mathbf{k})$ [Fig. \ref{fig:bs} (d)].  In comparison, $max(|\Delta_{\text{so}}^{4,5}|)=59\,$meV. 
The weaker $k$-dependence of $\Delta_{\text{so}}^{4,5}$ results from strong isolation of the bands 4 and 5, by $\sim 1.3\,eV$  from lower and upper manifolds (not shown), effectively limiting the possible couplings mainly to those two partners. 

\begin{table}
\caption{\label{tab:splittings} Spin-orbital energy shifts $\Delta_{\text{so}}^{\text{n,m}}$ at high symmetry points extracted from first principles calculations.}
\begin{ruledtabular}
\begin{tabular}{cccc}
$k$-point & $\Delta_{\text{so}}^{\text{1,2}}$ [meV] & $\Delta_{\text{so}}^{\text{4,5}}$ [meV] & $\Delta_{\text{so}}^{\text{2,3}}$ [meV]  \\
\hline 
$\Gamma$ & 350  &  59 & -4\\
K & 170 & 5  & -24
\end{tabular}
\end{ruledtabular}
\end{table}

\textit{Spin mixing}.
Apart from  the spectroscopic features discussed above, the strength of SOC of inversion-symmetric crystals is measured by the spin mixing parameter $b^2_{\mathbf{k}}$. Because $b^2_{\mathbf{k}}$ originates from the intrinsic SOC it constitutes a good measure of this interaction in the band structure \citep{Kurpas2019}. Exceptions are spin \textit{hot spots} formed around high-symmetry and accidental degeneracy points at which the value of $b^2_{\mathbf{k}}$ is strongly enhanced \citep{fabian1998,Kurpas2019} and the mixing reaches the value of one half (equal probability for spin up and down in a given state), irrespective of the strength of SOC.

For an arbitrary Bloch state 
\begin{eqnarray}
\Psi^\Uparrow_{n,{\mathbf k}}({\mathbf r}) & = & \left[ a_{n,{\mathbf k}}({\mathbf r})\vert \uparrow \rangle + b_{n,{\mathbf k}}({\mathbf r})\vert \downarrow \rangle \right] e^{i {\mathbf k}\cdot {\mathbf r}},
\label{eq:spinor}
\end{eqnarray} 
where $n$ is the band index, $a_{n,{\mathbf k}}$ and $b_{n,{\mathbf k}}$ are lattice periodic functions, $|\sigma\rangle$, $\sigma=\lbrace\uparrow,\downarrow\rbrace$ is an eigenstate of spin one-half operator and $\mathbf{k}$ is the crystal momentum, the spin mixing parameter is defined as
\begin{equation}
    b^2_{\mathbf{k}} = \int \vert b_{n,{\mathbf k}}({\mathbf r})|^2 d\mathbf{r},
\end{equation}where the integral is taken over the entire unit cell. Here, the amplitudes $a_{n,{\mathbf k}}({\mathbf r})$ and $b_{n,{\mathbf k}}({\mathbf r})$ are chosen in a way, that $b_{n,{\mathbf k}}({\mathbf r})$ is the amplitude of the spin component admixed by the SOC. Such a choice is possible for any spin quantization axis (SQA).  
Because usually $ |b_{n,{\mathbf k}}({\mathbf r})| \ll |a_{n,{\mathbf k}}({\mathbf r})|$  the state (\ref{eq:spinor}) can still be called a \textit{spin up} state (although it is not an eigenstate of a Pauli matrix) \citep{Zutic2004:RMP}. For centrosymmetric systems with time reversal symmetry the energy degenerate \textit{spin down} partner of $\Psi^\Uparrow_{n,{\mathbf k}}({\mathbf r})$ is 
\begin{eqnarray}
\Psi^\Downarrow_{n,{\mathbf k}}({\mathbf r}) & =&  \left[ a^*_{n,-{\mathbf k}}({\mathbf r})\vert \downarrow \rangle - b^*_{n,-{\mathbf k}}({\mathbf r})\vert \uparrow \rangle \right] e^{i {\mathbf k}\cdot {\mathbf r}},
\label{eq:spinor2}
\end{eqnarray}
and the same definition of $b^2_{\mathbf{k}}$ can be used. It is immediately seen that for normalized states $b^2_{\mathbf{k}}\in [0;0.5]$, where $b^2_{\mathbf{k}}=0$ means no spin mixing and $b^2_{\mathbf{k}}=0.5$ for fully spin mixed states. Alternatively, $b^2_{\mathbf{k}}$ can be defined as a deviation of the spin expectation value from one half \citep{zimmermann2012}.\\

\begin{figure}
    \centering
    \includegraphics[width=0.98\columnwidth]{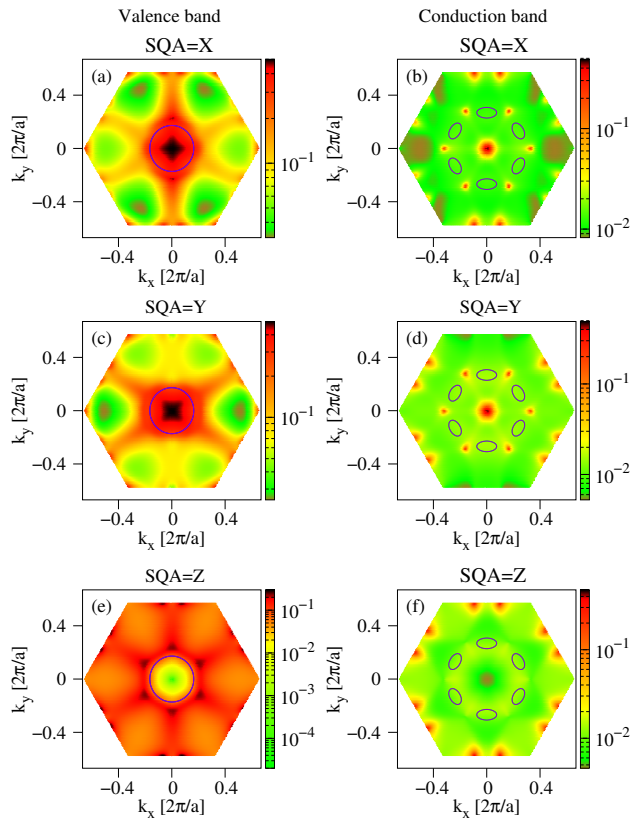}
    \caption{\label{fig:mb2k_map}  Distribution of the spin-mixing parameter $b^2_{k}$ in the first Brillouin zone of PtSe$_2$ for three spin quantization axes. Left column: (a) valence band and SQA=X, (c) valence band and SQA=Y, (e) valence band and SQA=Z.  Right column: same as left but for the conduction band. Blue s and ellipses encircle wedges of the BZ corresponding to the carried density up to  $n=12\cdot 10^{13}$\,cm${^-2}$.}
\end{figure}

To quantify  anisotropies in the spin relaxation and spin transport in the crystal, it is instructive to study the spin admixture parameter for different spin quantization axes, which correspond to either the direction of an applied magnetic field or to the orientation of the injected spin in a spin injection experiment. 
In Fig. \ref{fig:mb2k_map} we show $b^2_{\mathbf{k}}$ calculated in the first Brillouin zone (FBZ) for the highest valence and first conduction bands, and for three different spin quantization axes SQA=$\lbrace$X,Y,Z$\rbrace$ aligned with the real space axes shown in Fig. \ref{fig:structure}. 
A strong anisotropy of $b^2$ is evident. In the valence band  and for SQA=X/Y [Fig. \ref{fig:mb2k_map} (a), (c)] the region around the BZ center is a spin hot region where $b^2_{\mathbf{k}}$ is close to one half. 
This region is very wide and extends towards the M-points, in a different way for SQA=X and SQA=Y. 
At the $\Gamma$ point spins are fully mixed,  $b^2_{\mathbf{k}}\approx 0.5$; this is a witness
to the lifting of the orbital degeneracy by SOC. 
For SQA=Z [Fig. \ref{fig:mb2k_map} (e)], the entire FBZ is a spin hot region with the value of  $b^2_{\mathbf{k}} \sim 10^{-2}-10^{-1}$. An exception is a small circular wedge in the center of BZ corresponding to the vicinity of the valence band maximum [see Fig. \ref{fig:bs} (a)]. In this wedge $b^2_{\mathbf{k}}$ varies from $10^{-5}$ in the center to $10^{-2}$ at the edge. \\
In the conduction band [Fig. \ref{fig:mb2k_map} (b), (d), (f)] we observe much smaller variation of spin mixing parameter than for the valence band. The value of $b^2_{\mathbf{k}}$ is of the order of $10^{-2}$ within the whole BZ, except for several spin hot spot regions localized around high symmetry and accidental degeneracy points. 

\begin{figure}
    \centering
    \includegraphics[width=0.99\columnwidth]{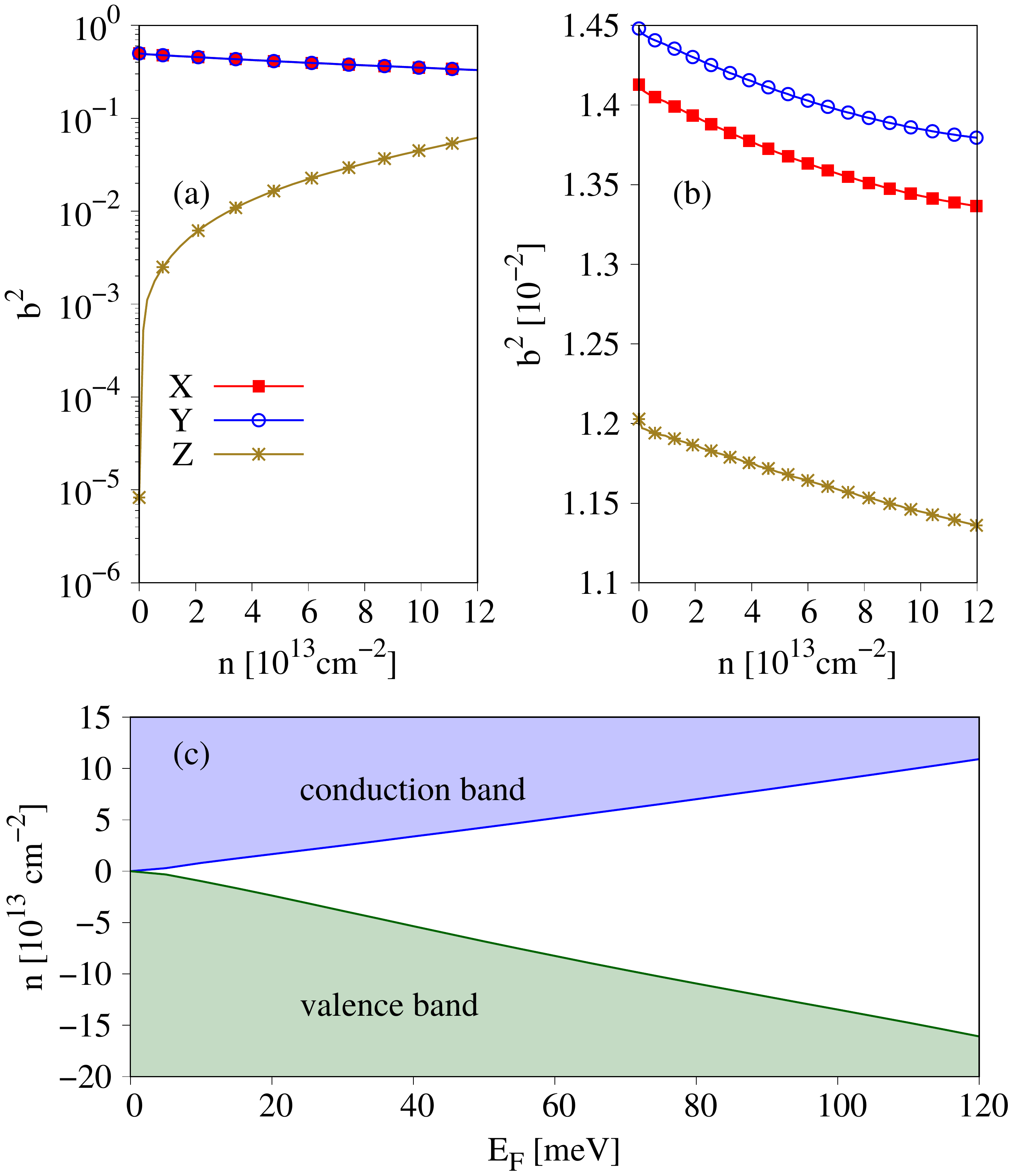}
    \caption{\label{fig:b2k}  Calculated Fermi surface averaged spin--mixing parameter $b^2$ versus carrier density $n$ for the valence (a) and for the conduction band (b). (c) Carrier density versus position of the Fermi level given with respect to valence band maximum and conduction band minimum. }
\end{figure}

According to Elliott \citep{elliott_theory_1954}, the spin mixing parameter can be translated into the spin relaxation rate, provided we know the momentum relaxation time. 
The latter strongly depends on temperature, concentration of defects and dopants and for a given sample can be determined from transport experiments.  Here, we calculate the intrinsic, sample independent property of PtSe$_2$ required to estimate spin lifetime -- the Fermi surface averaged spin mixing parameter $b^2$. It is shown in Fig.  \ref{fig:b2k} (a),(b) as a function of carrier density $n$, plotted versus Fermi energy in Fig.  \ref{fig:b2k} (c).
As can be seen,  $b^2$ in the valence band displays a qualitatively different behaviour for in-plane and out-of-plane SQA [Fig.  \ref{fig:b2k} (a)].
For SQA=Z it is growing exponentially from $b^2\approx 10^{-5}$ to $b^2=10^{-1}$ when the hole density is increasing. For in-plane spin polarization  $b^2$ slowly decreases from the value of about 0.5 with increasing $n$, but never gets below $10^{-1}$. 
This unusually large value of $b^2$ is due to the very broad spin hot region. In this case  the perturbative Elliott's approach to $\tau_s$ is not valid and the spin relaxation rate is essentially the same
as momentum relaxation.
Therefore, if the momentum relaxation anisotropy (in the plane) is not very large, one should expect a giant, doping dependent anisotropy of the spin relaxation in PtSe$_2$ for holes.

In contrast, the spin relaxation anisotropy is predicted
to be rather weak for conduction electrons. Indeed, 
as seen in Fig. \ref{fig:b2k} (b), $b^2$ in the conduction band varies very little with $n$. Its value is of the order of $10^{-2}$, meaning that out of all momentum 
scattering events about 1\% constitute a spin flip. 
Moreover, there is a very weak dependence of the spin mixing probability on SQAs. The spin relaxation rate for out-of-plane spins is expected to be somewhat slower than for in-plane spins. Also, spin lifetimes of electrons should
be 1-2 orders longer than for holes, for in-plane spins.
\subsection{Extrinsic spin-orbit coupling.} In realistic situations monolayers are often encapsulated in protective layers, sit on a substrate or are studied in a gating electric field. In any of these configurations the space inversion symmetry is broken, leading to the symmetry reduction $D_{3d}\rightarrow C_{3v}$ in the case of monolayer PtSe$_2$; spin degeneracy $\varepsilon_{\mathbf{k},\uparrow} = \varepsilon_{\mathbf{k},\downarrow}$ is lifted, except at the time reversal points $\Gamma$ and $M$. 
The emerging spin-orbit fields $\Omega_{\mathbf{k}}$ enable  the D'yakonov-Perel mechanism of spin relaxation, which coexists with the Elliott-Yafet spin-flip scattering mechanism. 
We model the effects of space inversion symmetry breaking by applying a uniform external electric field $E$ in the direction perpendicular to the PtSe$_2$ sheet. In this approach the spin-orbit field depends on both the
momentum and electric field, $\Omega_{\mathbf{k}}(E)$, and is related to the spin splitting as
\begin{equation}
    H_{soc}= \frac{\hbar}{2}\Omega_{\mathbf{k}}(E)\cdot \boldsymbol{\sigma},
    \label{eq_Hex}
\end{equation}
 where $\hbar$ is the Planck constant, and $\boldsymbol{\sigma}$ is the vector of Pauli matrices.\\

\begin{figure}
    \centering
    \includegraphics[width=\columnwidth]{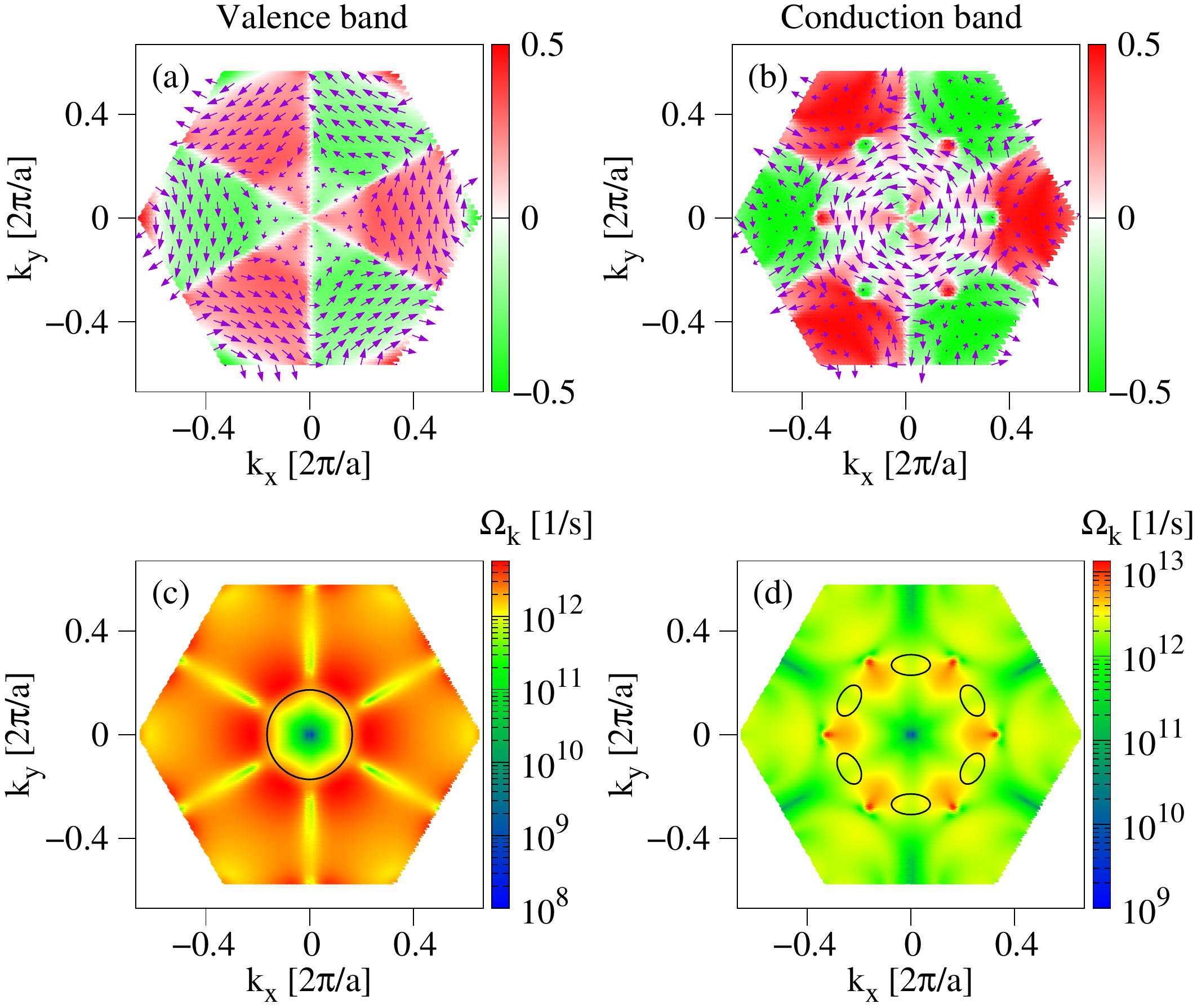}
    \caption{\label{fig:omega_fbz} Extrinsic SOC in PtSe$_2$: (a) In-plane spin texture (arrows) and out-of-plane  spin component  (color) of one spin subband of the top-most valence band plotted in the whole FBZ. (b) Same as (a) but for the bottom most conduction band. (c) The distribution of the spin-orbit field $\Omega_{\bf{k}}$ in the FBZ for the valence band for the electric field $E=1\,$V/nm. (d) Same as (c) but for the conduction band. Black circle and ellipses encircle wedges of the BZ corresponding to the carried density up to  $n=12\cdot 10^{13}$\,cm$^{-2}$.
    }
\end{figure}

In Fig. \ref{fig:omega_fbz} (a), (b) we show spin textures of the upper spin split valence band and for the lower spin split conduction band respectively and for the external electric field E=1\,V/nm.  The in-plane spin components (arrows) display a Rashba-like helical pattern, while the out-of-plane components (color) show the spin--valley locking effect. Similar spin textures have been reported to exist as a result of hidden spin polarization of spin degenerate bands in  PtSe$_2$ \citep{Yao2017}. In contrast to layer-resolved spin textures picturing local Rashba effect \citep{Yao2017}, Fig. \ref{fig:omega_fbz} (a), (b) shows a global spin texture of the entire crystal structure. To answer the question about the origin of the presented spin textures we performed calculations at zero electric field in order to preserve spin degeneracy of bands but with broken space inversion symmetry. The obtained spin textures (not shown here) resemble the same features as those shown in Fig. \ref{fig:omega_fbz} (a), (b), with small differences due to external electric field in the latter case.  
This indicates that such a texture is intrinsic to PtSe$_2$ crystalline structure and appears immediately once the space inversion symmetry is broken by arbitrary small crystal potential imbalance.

\begin{figure}
    \centering
    \includegraphics[width=0.99\columnwidth]{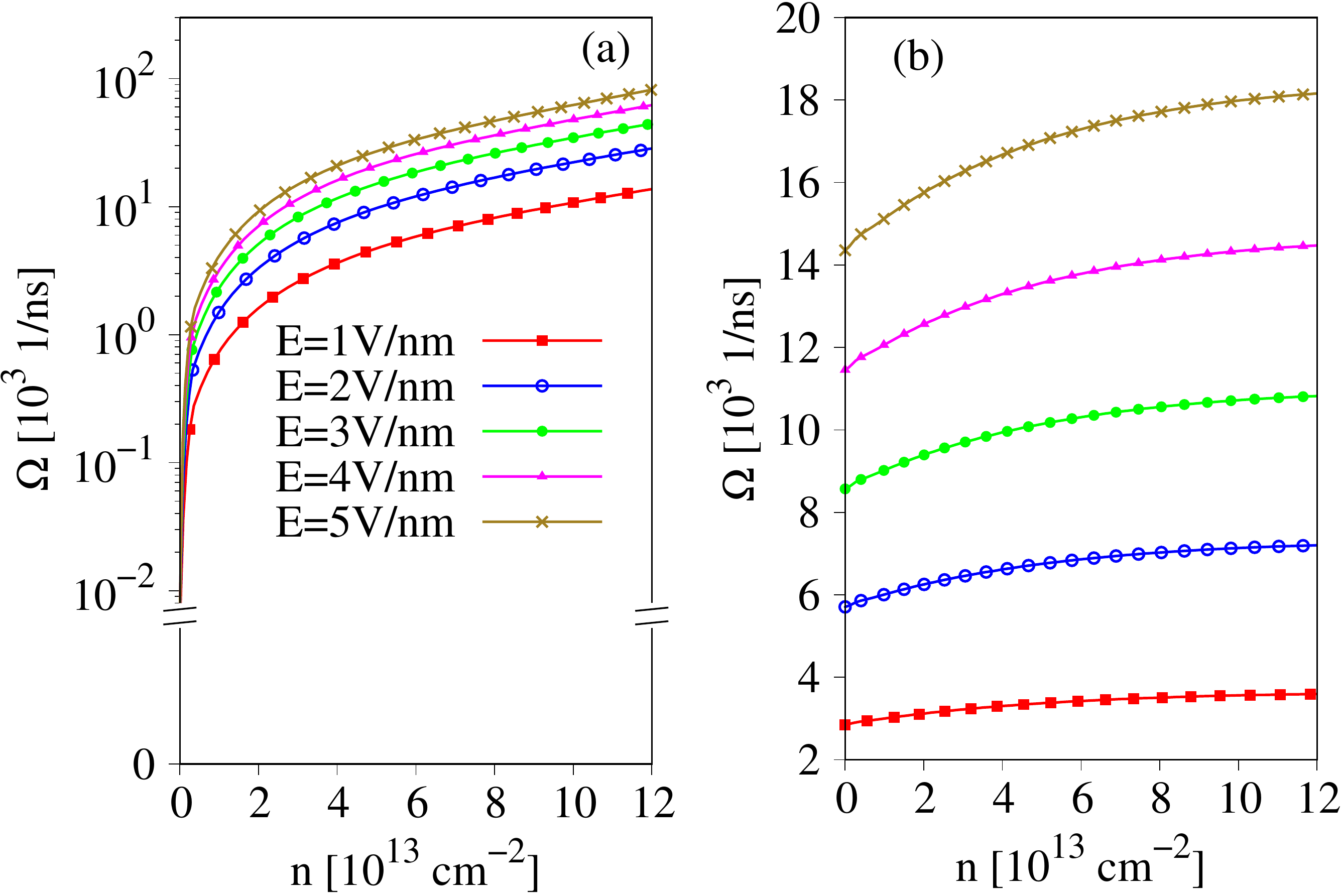}
    \caption{\label{fig:omega}  Calculated Fermi surface averaged spin--orbital field $\Omega$ versus carrier density $n$ and several values of electric field perpendicular to the 2D lattice. (a) valence band, (b) conduction band.}
\end{figure}

Let us now discuss the strength of the extrinsic SOC. It can be quantified by the amplitude of the spin-orbital fields $\Omega_{\mathbf{k}}$. In Figs. \ref{fig:omega_fbz} (c),(d) and \ref{fig:omega}  we show the $\mathbf{k}$-point resolved ($\Omega_{\mathbf{k}}$) and Fermi surface averaged ($\Omega$) spin-orbit fields, respectively,
for the electric field of 1 V/nm.
In the valence band the overall value of $\Omega_\mathbf{k}$ in the FBZ is greater that in the conduction band. This can be seen by comparing  the amount of the red shaded area in Fig. \ref{fig:omega_fbz} (a) and (b). A very characteristic hexagonal structure is formed close to the BZ center, with corners pointing towards the M points. In this region, and also along the paths $\Gamma$M and around the K points, the spin splitting for $E=1$\,V/nm is less than 4\,meV (the maximal value in the valence band and $E=1$\,V/nm is 14\,meV). This gives $\Omega_{\mathbf{k}} \sim 10^4$\,ns$^{-1}$, which is roughly the value of $\Omega$ for $n$ between $8\cdot 10^{13}$cm$^{-2}$ and $12\cdot10^{13}$cm$^{-2}$ [see the red line in Fig. \ref{fig:omega} (a)]. 
Within the full doping range (without the first 5\,meV), $\Omega$ in the VB varies by an order of magnitude.\\

In the conduction band the extrinsic SOC is weaker than in the VB, resulting in smaller values of $\Omega_{\mathbf{k}}$ [Fig. \ref{fig:omega_fbz} (b)]. For the same level of doping, e.g., $n=10^{14}$cm$^{-2}$, 
the ratio of $\Omega$ in the VB to  $\Omega$ in the CB, $\Omega_{VB}/\Omega_{CB}\approx 3$.
 Similarly to $b^2$, $\Omega$ in the conduction band is very weakly doping dependent [Fig. \ref{fig:omega} (b)].
For E=1\,V/nm it is of the order of $10^3$\,ns$^{-1}$, and grows approximately with a step $3\cdot 10^3$ per 1\,V/nm.

\subsection{Spin lifetime.} 
To estimate spin lifetimes $\tau_{s,\rm EY}$ and $\tau_{s,\rm DP}$ we need to know the momentum relaxation time $\tau_p$. Taking the experimental value of the mobility for electrons  $\mu=400$\,cm$^2$V$^{-1}$s$^{-1}$ (100K) and effective mass $m^*=0.37 m_e$ \citep{Zhao2017}, we estimate, using the Drude formula $\tau_p = \mu e/m^*$, $\tau_p\approx 80$\,fs. 
The formula (\ref{eq:elliott}) for the Elliott-Yafet spin relaxation rate is valid under the assumption that $b^2$ can be treated as a small parameter, $b^2 \ll 1$ \citep{elliott_theory_1954}. For valence electrons being polarized in-plane with the PtSe$_2$ sheet (SQA=X/Y), $b^2$ is of the order of 1 [Fig. \ref{fig:b2k}], and thus Eq. (\ref{eq:elliott}) cannot be applied. In such a case, due to very strong spin mixing,  spin lifetime should be limited by the momentum relaxation  time, i.e., $\tau_{s,\rm EY} \approx \tau_p \approx 80$\,fs.
For spins of valence electrons being polarized out-of-plane (SQA=Z) and for conduction electrons, $b^2 \approx 10^{-2}$, within the perturbative limit. The corresponding spin lifetime estimated from Eq. (\ref{eq:elliott}) is $\tau_{s,\rm EY} \approx 1$\,ps.

Two qualitatively different regimes of spin relaxation apply also for the D'yakonov-Perel mechanism. For small fields $E\le 1\,$V/nm, we are in the motional narrowing regime, i.e., $\tau_p \Omega \ll 1$ \citep{Zutic2004:RMP}. Taking $\Omega=3\cdot 10^3\,$ns$^{-1}$ (E=1\,V/nm) we get for out-of-plane spins  $\tau_{s,\rm DP} \approx \tau_{s,\rm EY} \approx 1\,$ps. 
The in-plane spins in pristine paramagnetic materials usually  are expected to relax slower than the out-of-plane ones due to 2-dimensional (in-plane) character of extrinsic Rashba spin-orbit fields. In the Rashba limit, the  in-plane spin relaxation rate is 2, which gives, roughly, the same order of magnitude as for out-of-plane spins.
With increasing  electric field the condition for motional narrowing breaks down ($\Omega \ge 10^4$\, ns$^{-1}$), and irreversible spin randomization occurs in the time scale given by the momentum relaxation, $\tau_{es,\rm DP}\approx \tau_p$  \citep{Zutic2004:RMP}, irrespective of spin polarization.

\section{Conclusions}\label{sec:conclusions}
We have investigated the intrinsic and extrinsic spin-orbit couplings, and their influence on the electronic properties spin relaxation in monolayer PtSe$_2$ using first principles calculations. We found that the intrinsic SOC is very strong and leads to a significant mixing of the spin states. The extrinsic SOC, characterized by spin-orbit fields $\Omega$, is also expected to be large, on the
order of the intrinsic one. Their interplay is manifested in comparable contributions of the Elliott-Yafet and D'yakonov-Perel mechanisms to spin relaxation. 
Spin lifetime in PtSe$_2$ is predicted to be short, on the picosecond time scale, and to a large extent governed by the momentum relaxation time especially at spin hot spots where the spins are fully mixed. This brings serious limitations for using  PtSe$_2$ in spintronic devices requiring long spin lifetimes. On the other hand, the strong spin-orbit coupling should be manifested prominently in spin transport and spin control phenomena. 

\section*{Acknowledgments}\label{sec:ack}
This work was supported by the National Science Centre under the contract DEC-2018/29/B/ST3/01892, in part by PAAD Infrastructure co-financed by Operational Programme Innovative Economy, Objective 2.3,
and by the Deutsche
Forschungsgemeinschaft (DFG, German Research Foundation) – Project-ID
314695032 – SFB 1277.

\section{References}
\bibliography{references.bib}
\end{document}